\documentclass[useAMS,usenatbib]{mn2e}
\usepackage{graphicx}

\title[Evolution of CoRoT-7b and Kepler-10b]{Atmospheric mass loss and evolution of short-period exoplanets: the examples of CoRoT-7b and Kepler-10b}
\author[H. Kurokawa and L. Kaltenegger]{H. Kurokawa$^{1,2}$\thanks{E-mail:
kurokawa@nagoya-u.jp} and L. Kaltenegger$^{2,3}$\\
$^{1}$Tokyo Institute of Technology, 2-12-1 Ookayama, Meguro- ku, Tokyo 152-8551, Japan\\
$^{2}$Max Planck Institut fuer Astronomie, Koenigstuhl 17, 69117, Heidelberg, Germany\\
$^{3}$Harvard Smithsonian Center for Astrophysics, 60 Garden St., 02138 MA,Cambridge, USA}
\begin{document}



\maketitle

\label{firstpage}

\begin{abstract}
Short-period exoplanets potentially lose envelope masses during their evolution because of atmospheric escape caused by the intense XUV radiation from their host stars. 
We develop a combined model of atmospheric mass loss calculation and thermal evolution calculation of a planet to simulate its evolution and explore the dependences on the formation history of the planet.
Thermal atmospheric escape as well as the Roche-lobe overflow contributes to mass loss.
The maximum initial planetary model mass depends primarily on the assumed evolution model of the stellar XUV luminosity.
We adapt the model to CoRoT-7b and Kepler-10b to explore the evolution of both planets and the maximum initial mass of these planets. 
We take the recent X-ray observation of CoRoT-7 into account and exploring the effect of different XUV evolution models on the planetary initial mass.
Our calculations indicate that both hot super Earths could be remnants of Jupiter mass gas planets.
\end{abstract}


\section{Introduction}

Hundreds of short-period exoplanets and candidates (KOIs) have been found by transit and radial velocity observations \citep{bor11,bat12,auv09}. 
Most of these planets are located closer to their host star  than any planet in our Solar System.
Short-period exo-planets might lose substantial mass during their evolution because of atmospheric escape caused by intense XUV radiation (X-ray and extreme UV) from their host stars \citep[see e.g.][]{lam09}. 

As extreme cases, we focus on the evolution of the super Earths, CoRoT-7b \citep{log09} and Kepler-10b \citep{bat11}, where both mass and radius are known.
CoRoT-7b has a mass of $7.38 \pm 0.34\ M_\oplus$ \citep{hat11} and a radius of $1.58 \pm 0.10\ R_\oplus$ \citep{bru10}.
Kepler-10b has a mass of $4.56^{+1.17}_{-1.29}\ M_\oplus$ and a radius of $1.416^{+0.033}_{-0.036}\ R_\oplus$ \citep{bat11}.
These measured masses and radii are consistent with Earth-like rock and iron composition \citep[see e.g.][]{wag11}. 
Both CoRoT-7b and Kepler-10b orbit a G-type star and have short-period orbit, $0.0172 \pm 0.00029\ {\rm AU}$ \citep{log09} and $0.01684 \pm 0.00032\ {\rm AU}$ \citep{bat11}, respectively, providing extremely hot environments.

The mass loss evolution of CoRoT-7b was explored in previous studies, 
with different results, two studies found that CoRoT-7b could be a remnant of a Jupiter-mass gas giant which experienced intense atmospheric mass loss \citep{val10, jac10}. 
Another study found that CoRoT-7b would need to have formed as an initial rocky planet \citep{lei11}.
The difference found in these evolution studies is a result of the different evolution models postulated for the planet as well as different stellar XUV luminosity evolution used, one based on the {\it Sun in Time} project \citep{rib05} used in \citet{val10} and \citet{jac10} versus an evolution model of soft X-ray luminosity \citep{pen08} as scale for the total XUV luminosity used in \citet{lei11}.
Recent observations of the X-ray luminosity of CoRoT-7 \citep{pop12} estimated the whole XUV luminosity to be $2.4 \times 10^{29} \ {\rm erg \ s^{-1}}$, larger by an order of magnitude than the values $5.1 \times 10^{28} \ {\rm erg \ s^{-1}}$ \citep{rib05} and $2.1 \times 10^{28} \ {\rm erg \ s^{-1}}$ \citep{pen08} estimated from XUV models for the assumed age of CoRoT-7 of $1.5\ {\rm Gyr}$.
Mass loss evolution of Kepler-10b indicates that Kepler-10b should have formed as a rocky planet \citep{lei11}. 
Their models were based on XUV model by \citet{pen08}.

In this paper, we explore the effect of different XUV models on mass loss history and evolution of a close-in planet.
We use CoRoT-7b and Kepler-10b as examples of two different close-in planetary environments, to investigate in which initial states both planets could have been as well as their evolution. 
Section 2 introduces our numerical models, section 3 presents, and section 4 discusses our results.

\section[]{Numerical Models}

Our planet evolution model solves for atmospheric mass loss and thermal contraction simultaneously.
Jupiter-like gas giant planets having solar composition envelopes on solid cores are assumed as an initial state.
The planetary radius and the intrinsic luminosity are obtained with a structure calculation (see \ref{structure}) for each time step. 
The thermal energy loss rate is calculated using the intrinsic luminosity and the atmospheric escape rate is calculated using the stellar XUV flux.
The incoming XUV flux is derived from the evolution of the XUV luminosity model of the host stars (see \ref{evolution}).

\subsection{Structure Calculation} \label{structure}

We assume a spherically symmetric planet with a solar composition envelope and a rocky core.
The rocky core is assumed to have the observed mass and radius values of the planet. 
The effect of the core radius evolution on the evolution of a planet is neglected.
The atmospheric envelope is in hydrostatic equilibrium and is divided into an upper radiative equilibrium layer and a lower convective layer.
The mass conservation equation and the hydrostatic equation (Eq. \ref{mass} and Eq. \ref{hydro}) are solved for the whole envelope, 
\begin{equation}
\frac{{\rm d} \ln{r}}{{\rm d} M_r} = - \frac{1}{4 \pi r^3 \rho},\label{mass}
\end{equation}
\begin{equation}
\frac{{\rm d} \ln{p}}{{\rm d} M_r} = - \frac{G M_r}{4 \pi r^4 p},\label{hydro}
\end{equation}
where, $r$ is the distance from the center of the planet, $M_r$ is the enclosed mass at the distance $r$, $p$ is the pressure, $\rho$ is the density, and $G$ is the gravitational constant.
To calculate the temperature structure, an analytic solution of the plane-parallel two-stream approximation \citep{gui10} is used in the radiative equilibrium layer (Eq. \ref{radiative}),
\begin{equation}
T^4 = \frac{3 T_{\rm int}^4}{4} \Bigl( \frac{2}{3} + \tau \Bigr)
           + \frac{3 T_{\rm irr}^4}{4} f \Bigl( \frac{2}{3} + \frac{1}{\gamma \sqrt{3}} + \bigl( \frac{\gamma}{\sqrt{3}} - \frac{1}{\gamma \sqrt{3}} \bigr) e^{-\gamma \tau \sqrt{3}} \Bigr),\label{radiative}
\end{equation}
where $T$ is the temperature, $\tau$ is the optical depth for outgoing long-wave radiation, $T_{\rm int}$ is the intrinsic temperature, $T_{\rm irr}$ is the irradiation temperature defined with the equilibrium temperature $T_{\rm eq} = f^{1/4} T_{\rm irr}$, $f$ is the redistribution factor ($f=1/4$ for full redestribution), and $\gamma$ is the ratio of the short-wave to long-wave optical depth.
We take $\gamma = 0.6 \sqrt{T_{\rm irr}/2000{\rm K}}$ \citep[following][]{gui10}, which provides a good match to detailed calculations of hot Jupiter atmospheres \citep{for08}.
The intrinsic temperature $T_{\rm int}$ is defined by the intrinsic luminosity $L_{\rm int} = 4 \pi R_{\rm p}^2 \sigma_{\rm SB} T_{\rm int}^4$, where $R_{\rm p}$ is the planetary radius and $\sigma_{\rm SB}$ is the Stefan-Boltzmann constant.
The optical depth $\tau$ is calculated using Eq. \ref{tau},
\begin{equation}
\frac{{\rm d} \ln{\tau}}{{\rm d} M_r} = - \frac{\kappa}{4 \pi r^2 \tau},\label{tau}
\end{equation}
where $\kappa$ is the opacity for long-wave radiation.
The plane-parallel approximation is only valid when the radiative equilibrium layer is thin compared to the planetary radius. 
This condition breaks down for highly irradiated atmospheres.
Therefore we use the diffusion approximation equation \citep{rog11} (Eq. \ref{diffuse}) for the lower part of the radiative equilibrium layer where incoming stellar radiation is not effective ($\tau \gg 1/\sqrt{3} \gamma$), 
\begin{equation}
\frac{{\rm d} T}{{\rm d} r} = - \frac{3 \kappa \rho}{16 \sigma_{\rm SB} T^3} \frac{L_{\rm int}}{4 \pi r^2}.\label{diffuse}
\end{equation}
In the lower convective layer, an adiabatic temperature structure (Eq. \ref{adiabatic}) is assumed,
\begin{equation}
\frac{{\rm d} \ln{T}}{{\rm d} M_r} = - \frac{G M_r}{4 \pi r^4 p} \Bigl( \frac{\partial \ln{T}}{\partial \ln{p}} \Bigr)_s,\label{adiabatic}
\end{equation}
where $S$ is the entropy.
The boundary between the radiative equilibrium layer and the convective equilibrium layer is determined by comparing the temperature lapse rate for a radiative equilibrium and an adiabatic lapse rate.
We use the data table of \citet{sau95} for the equation of state of solar composition gas and the Rosseland mean opacity data table of \citet{fre08} for gas opacity. 
Out of the range of the opacity table, we use an power-law dependance Eq. \ref{power} \citep[following][]{rog10},
\begin{equation}
\kappa = C p^\alpha T^\beta,\label{power}
\end{equation}
where $\log C = -7.32$, $\alpha = 0.68$, $\beta = 0.45$ with all quantities in SI units.
The upper boundary is defined as $R_{\rm p}$ at the layer where $\tau=2/3$ and the pressure at the layer is obtained by Eq. \ref{upper_boundary},
\begin{equation}
\kappa p = \frac{2}{3} g,\label{upper_boundary}
\end{equation}
where $g$ is the gravity at the layer \citep[following][]{rog11}. 
The lower boundary is set to the radius of the core $R_{\rm core}$.
The equations are solved iteratively until a self-consistent structure for the assumed set of planetary mass $M_{\rm p}$, core mass $M_{\rm core}$, core radius $R_{\rm core}$, irradiation temperature $T_{\rm irr}$, and intrinsic temperature $T_{\rm int}$ is obtained.
The observable planetary radius $R_{\rm p}$ is obtained as a result.

\subsection{Evolution Calculation} \label{evolution}

Thermal evolution is calculated by integrating the energy equation (Eq. \ref{evo_thermal}),
\begin{equation}
\int_{M_{\rm core}}^{M_p} T \frac{{\rm d} S}{{\rm d} t} {\rm d} M_r = - L_{\rm int}. \label{evo_thermal}
\end{equation}
Here we neglect the contribution of the core because its thermal energy and radioactive energy is small compared to the energy of a massive envelope for giant planets.
To integrate thermal evolution, entropy is assumed to be constant in the whole envelope, and to be the same as the entropy of the convective layer for simplification \citep[see e.g.][]{val10,lop12}.
The thermal atmospheric mass loss is calculated by integrating the energy-limited escape rate formula (Eq. \ref{evo_escape}) \citep{lop12},
\begin{equation}
\frac{{\rm d} M_{\rm p}}{{\rm d} t} = \frac{\eta \pi F_{\rm XUV} R_{\rm XUV}^3}{G M_{\rm p} K_{\rm tide}}, \label{evo_escape}
\end{equation}
where $\eta$ is the heating efficiency for which we use $25\ \%$ as the biggest estimate of mass loss \citep{lei11}, $F_{\rm XUV}$ is the XUV flux, $R_{\rm XUV}$ is the distance from the center where the optical depth for XUV radiation equals unity, and $K_{\rm tide}$ is the correction factor that accounts for tidal effects in the Roche potential of planets \citep{erk07} given by,
\begin{equation}
K_{\rm tide} = \Bigl( 1-\frac{3}{2 \xi}+\frac{1}{2 \xi^3} \Bigr),
\end{equation}
\begin{equation}
\xi = \frac{R_{\rm rl}}{R_{\rm XUV}}.
\end{equation}
The Roche-lobe radius, $R_{\rm rl}$, is given by,
\begin{equation}
R_{\rm rl} = a_{\rm p} \Bigl( \frac{M_{\rm p}}{3 M_{\rm star}} \Bigr)^{\frac{1}{3}},
\end{equation}
where $a_{\rm p}$ is the orbital distance from the host star and $M_{\rm star}$ is the mass of the star.
$R_{\rm XUV}$ is calculated using Eq. \ref{RXUV}  \citep{rog11},
\begin{equation}
R_{\rm XUV} = R_{\rm p} + H_{\rm R} \ln{\frac{P_{\rm R} R_{\rm XUV}^2}{N_{\rm H} m_{\rm H} G M_{\rm p}}},\label{RXUV}
\end{equation}
where, $H_{\rm R}$ is the scale height at $r=R_{\rm p}$, $P_{\rm R}$ is the pressure at $r=R_{\rm p}$, $N_{\rm H} \sim 10^{17} \ {\rm cm^{-2}}$ is the column density of neutral hydrogen needed for the optical depth for XUV to equal unity, and $m_{\rm H}$ is the weight of a hydrogen atom.

The hydrodynamic wind is launched from the radius where the XUV radiation is absorbed. 
Therefore, we follow the model of \citet{lop12} and use $R_{\rm XUV}^3$ in the energy-limited escape rate formula (Eq. \ref{evo_escape}).
\citet{erk07} used $R_{\rm XUV} \times R_{\rm p}^2$ instead of $R_{\rm XUV}^3$.
\citet{val10} and \citet{lei11} used $R_{\rm p}^3$ instead of $R_{\rm XUV}^3$.
To estimate the sensitivity of the results to this parameter, 
we calculate the difference of the escape rate as $40\ \%$ by setting $(R_{\rm XUV}/R_{\rm p})^3 \sim (17/15)^3 \sim 1.4$ (from our results in Fig. \ref{RplRxuv}). 

By integrating Eq. \ref{evo_thermal} and Eq. \ref{evo_escape}, we obtain $M_{\rm p}$ and the entropy of the convective layer for the next time step.
Then a consistent structure is derived by solving the structure equations with $T_{\rm int}$ as input parameter.

\begin{figure}
  \includegraphics[width=0.99\columnwidth]{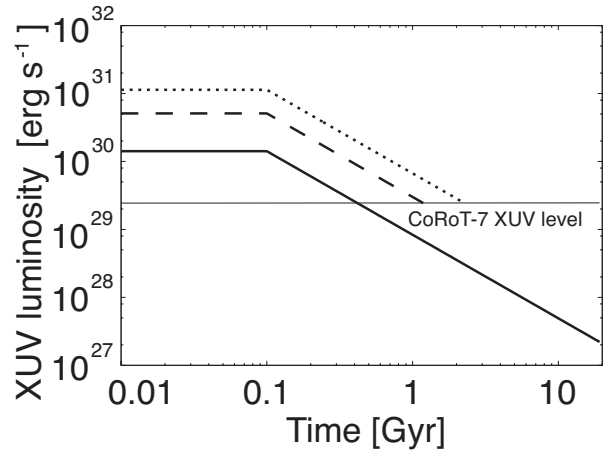}
  \caption{Evolution models of stellar XUV luminosity assumed in this work as a function of time. The solid line shows the scaling law of \citet{rib05}, assuming a saturation phase for the first $0.1\ {\rm Gyr}$. The dashed and dotted lines show a similar model scaled to two different XUV flux observations for CoRoT-7 \citep{pop12}. \label{XUV}}
\end{figure}

The evolution model of stellar XUV luminosity which we use in this study dominates the mass loss calculation (Fig. \ref{XUV}) for close-in planets.
Recent observations of the X-ray luminosity of CoRoT-7 \citep{pop12} estimated a total XUV luminosity of $L_{\rm XUV} = 2.4 \times 10^{29} \ {\rm erg \ s^{-1}}$ for the star. For Kepler-10 there are currently no XUV observations available. 
Based on the observation of stellar XUV luminosity of G-type stars at different ages, \citet{rib05} derived a scaling law (Eq. \ref{ribas}),
\begin{equation}
F_{\rm XUV} = 29.7 \Bigl(\frac{t}{1 {\rm Gyr}} \Bigr)^{-1.23} \ {\rm erg \ cm^{-2} \ s^{-1} \ \ at \ 1AU},\label{ribas}
\end{equation}
where $t$ is time.
We use the XUV luminosity model in eq. \ref{ribas} for Kepler-10.
For CoRot-7, we assume the same power law but fix the absolute value to the measurement \citep{pop12}.
Because stellar XUV level is known to be constant for young star \citep[e.g.][]{gar11}, called the \textquotedblleft saturation phase," we assume in our models that XUV luminosity is constant for the first $0.1\ {\rm Gyr}$ \citep[following][]{val10,lei11}.
The sensitivity of the results to other XUV models are tested (see Discussion).

The thermal atmospheric escape formula (Eq. \ref{evo_escape}) assumes that the Roche-lobe radius $R_{\rm rl}$ is larger than the XUV radius $R_{\rm XUV}$.
When $R_{\rm rl}$ becomes smaller than $R_{\rm XUV}$, the upper atmosphere is no longer bound by planetary gravity.
In that case, intense mass loss occurs until $R_{\rm rl}$ becomes larger than $R_{\rm XUV}$.
This mass loss mechanism is called {\it Roche-lobe overflow} and can be caused by orbital evolution \citep{gu03,jac10}.
In our models $R_{\rm XUV}$ becomes bigger than $R_{\rm pl}$ due to thermal atmospheric mass loss.
When $R_{\rm rl}$ becomes smaller than $R_{\rm XUV}$, we assume that the planet loses envelope mass rapidly until $R_{\rm rl}$ becomes larger than $R_{\rm XUV}$ without the thermal evolution by fixing the entropy of the convective layer.
The assumption is justified because the timescale of the overflow is $\sim 0.1 \ {\rm Myr}$ \citep{gu03} and is shorter than the time scale of thermal evolution and the age of the planets. 
Precise modeling of the overflow is beyond the scope of this paper.

\section{Results}

Our assumption on the initial state of planets follows \citet{val10}.
To mimic the phase of planet formation and dissipation of the proto-planetary disk, planets are assumed to cool without atmospheric mass loss for the first $10\ {\rm Myr}$.
The lifetime of the disks is known to have diversity and $10\ {\rm Myr}$ is a longer estimate \citep{Sicilia-Aguilar+2006} that provides a conservative value for mass loss.
We explore two formation scenarios, in situ formation and migration. 
In the migration scenario, planets are assumed to form at $0.1\ {\rm AU}$, and migrate to their current orbit after the first $10\ {\rm Myr}$.
We neglect the period of the migration and thus planets migrate to their current orbit at the time of $10\ {\rm Myr}$, conserving their entropy.
We take the initial intrinsic luminosity $L_{\rm int} = 10^{-5}\ L_\odot$, as a typical value for young gas giant after the formation phase \citep{mar07}.
The effect of the initial luminosity on the mass loss evolution is discussed (see Discussion).

\subsection{CoRoT-7b}

For CoRoT-7b, we assume a core mass of $M_{\rm core} = 7.38\ M_\oplus$, a core radius of $R_{\rm core} = 1.58\ R_\oplus$, and an orbital distance of $a_{\rm p} = 0.0172\ {\rm AU}$ in accordance with the observed values.
The effective temperature of CoRoT-7 is set to $5275\ {\rm K}$ \citep{log09}.
The age influences the stellar XUV flux model because its absolute current value is fixed to the measured value \citep{pop12}.
Because the age of CoRoT-7 is uncertain within $1.2\ {\rm Gyr}$ to $2.3\ {\rm Gyr}$ \citep{log09}, we model both values.

\begin{figure}
\includegraphics[width=0.99\columnwidth]{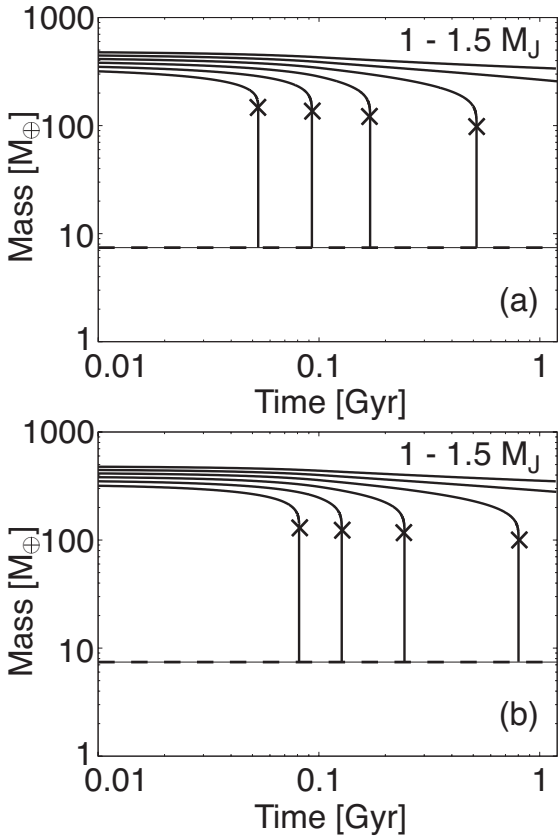}
\caption{Evolution of the mass of hypothetical CoRoT-7b precursors as a function of time for an age of ${\rm 1.2 \ Gyr}$ and a core mass according to CoRoT-7b's 7.38 Earth masses indicated by the dashed line and initial total masses from $1.0\ M_{\rm J}$ to $1.5\ M_{\rm J}$ in steps of $0.1\ M_{\rm J}$. Crosses indicate the time when Roche-lobe overflow occurs. Each panel shows the results of (a) in situ formation scenario, and, (b) migration scenario.\label{c7b_1-2Gy}}
\end{figure}

\begin{figure}
\includegraphics[width=0.94\columnwidth]{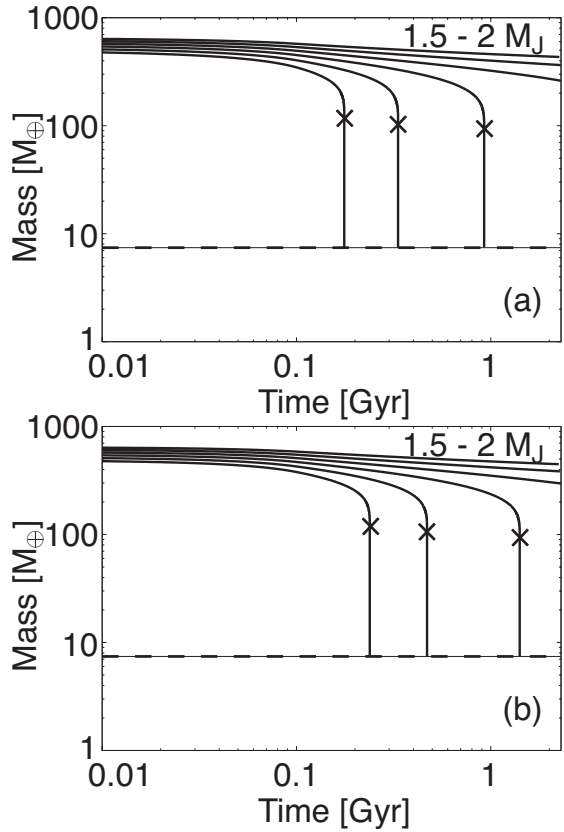}
\caption{Evolution of the mass of hypothetical CoRoT-7b precursors as a function of time for an age of ${\rm 2.3 \ Gyr}$ and a core mass according to CoRoT-7b's 7.38 Earth masses indicated by the dashed line and initial total masses from $1.5\ M_{\rm J}$ to $2.0\ M_{\rm J}$ in steps of $0.1\ M_{\rm J}$. Crosses indicate the time when Roche-lobe overflow occurs. Each panel shows the results of (a) in situ formation scenario, and, (b) migration scenario. \label{c7b_2-3Gy}}
\end{figure}

The evolution of the mass of hypothetical CoRoT-7b precursors is shown in Figs. {\ref{c7b_1-2Gy}} and \ref{c7b_2-3Gy} which correspond to the results of the two different stellar ages.
The maximum precursor masses which are consistent with CoRoT-7b's current mass are $1.3\ M_{\rm J}$ for a $1.2\ {\rm Gyr}$ stellar age model and $1.7\ M_{\rm J}$ for a $2.3\ {\rm Gyr}$ model.

\begin{figure}
\includegraphics[width=0.99\columnwidth]{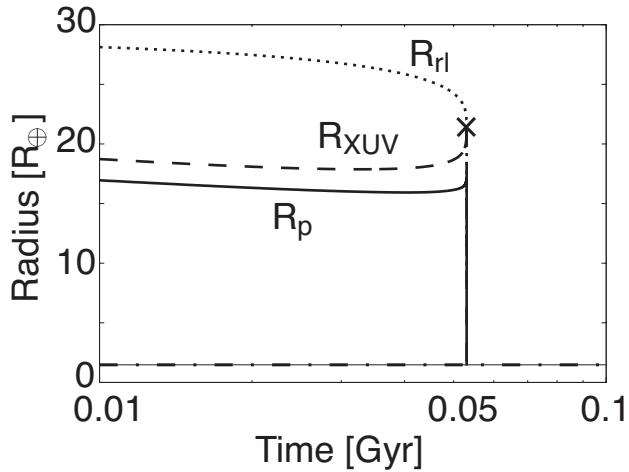}
\caption{Example of the evolution of the planetary radius (solid line), the XUV radius (dashed line), and the Roche-lobe radius (dotted line) as a function of time, for the model of a CoRoT-7b precursor assuming in situ formation with an initial mass of $1.0\ M_{\rm J}$ and an age of $1.2\ {\rm Gyr}$. The dash-dotted line is the assumed core radius. Crosses indicate the time when Roche-lobe overflow occurs. \label{RplRxuv}}
\end{figure}

In our models, Jupiter-mass precursors of both CoRoT-7b and Kepler-10b experience Roche-lobe overflow (shown as crosses in the figures) because of their small orbital distances from the host star.
Fig. \ref{RplRxuv} shows a typical example of the evolution of the planetary radius $R_{\rm p}$, the XUV radius $R_{\rm XUV}$, and the Roche-lobe radius $R_{\rm rl}$.
The mass loss caused by the thermal atmospheric escape decreases the Roche-lobe radius over the planet's evolution as well as increases the planetary radius before the overflow phase because of the decrease of planetary gravity.
After the Roche-lobe overflow, the planet maintains only $\sim 0.01\ M_\oplus$ atmospheric mass and can loose the remaining atmosphere in a short time scale.

Slight differences of the initial mass result in large differences of the final planetary mass.
This is because the thermal mass loss of the gas envelope decreases the density of the planets.
It results in the increase of the mass loss rate (see Eq. \ref{evo_escape}).
Roche-lobe overflow reduces the mass of the gas envelope even stronger.
As a result, a difference of $0.1\ M_{\rm J}$ in initial mass can change the planet's evolutionary model from a super Earth which lost its whole envelope to a sub Jupiter planet which still maintains a massive envelope.

\begin{figure}
  \includegraphics[width=0.99\columnwidth]{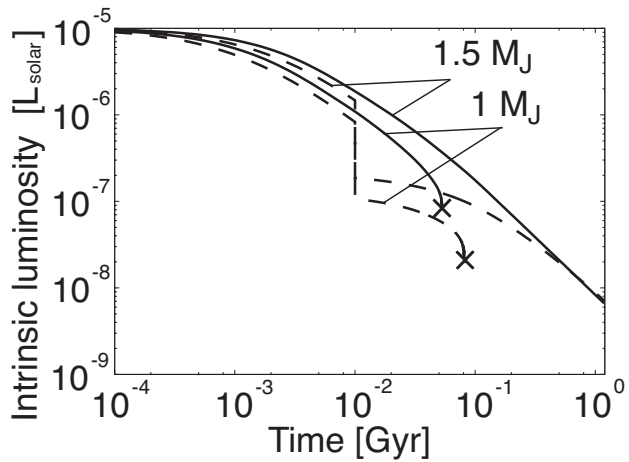}
\caption{Examples of the evolution of the intrinsic luminosity for models of CoRoT-7b for an age of $1.2\ {\rm Gyr}$. The solid lines and the dashed lines indicate the results for the case of in situ formation and migration respectively. The lower lines of the solid lines and the dotted lines which end at the Roche-lobe overflow points are the results of $1.0\ M_{\rm J}$. The upper lines are the results of $1.5\ M_{\rm J}$. \label{Lint}}
\end{figure}

The different estimates of the age of CoRoT-7 changes the critical mass that can lose its whole envelope by $\sim 0.4\ M_{\rm J}$ because it changes the total value of the XUV luminosity received over the planet's evolution.
The effect of the difference of the formation scenarios and respective distance where the planet initially formed on the critical mass is rather small $< 0.1\ M_{\rm J}$ (not visible in Figs. {\ref{c7b_1-2Gy}} and \ref{c7b_2-3Gy} that use a $0.1\ M_{\rm J}$ grid).
This slight difference is caused by the difference in the initial radius of the planet.
Migrated planets in our model have a smaller initial radius than planets which formed in situ, because a planet cools more efficiently when it is located in a cooler environment.
The difference of the thermal states is shown by the intrinsic luminosities in Fig. \ref{Lint}.

\subsection{Kepler-10b}

For Kepler-10b, we assume a core mass of $M_{\rm core} = 4.56\ M_\oplus$, a core radius of $R_{\rm core} = 1.416\ R_\oplus$, and an orbital distance of $a_{\rm p} = 0.01684\ {\rm AU}$, in accordance with observed values \citep{bat11}.
The effective temperature of the host star Kepler-10 is set to $5627\ {\rm K}$.
The estimated age of Kepler-10 is $11.9 \pm 4.5\ {\rm Gyr}$.
We calculate the evolution of the hypothetical Kepler-10 precursors using a standard XUV luminosity model \citep{rib05}. 
Therefore, the uncertainty of the age affects the total value of XUV received through the planet's evolution due to the lifetimes of the star but does not affect the absolute value of the XUV flux at each point of its evolution as in the case of CoRoT-7b.

\begin{figure}
  \includegraphics[width=0.99\columnwidth]{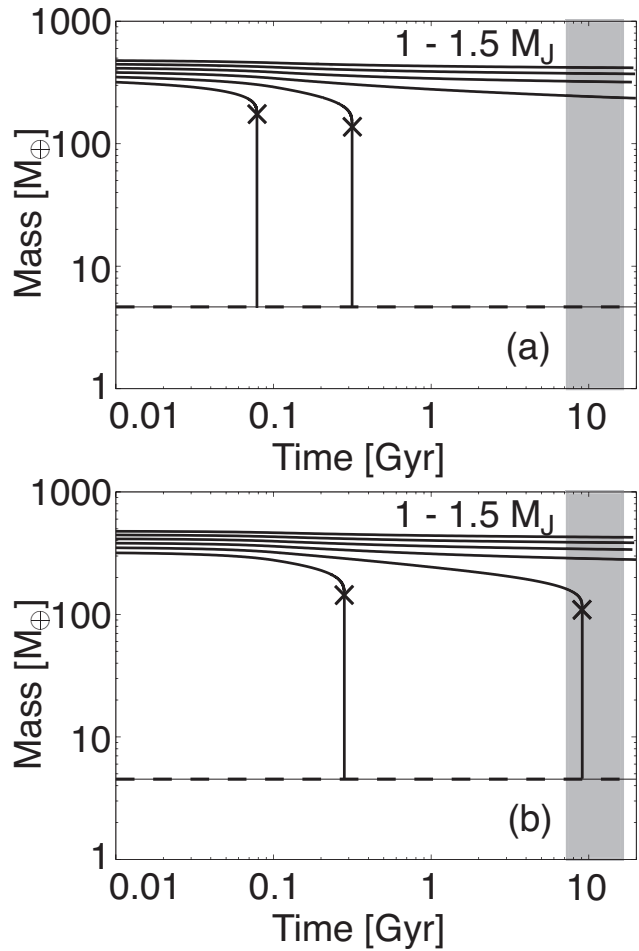}
\caption{Evolution of the mass of hypothetical Kepler-10b precursors as a function of time for a core mass according to Kepler-10b's 4.56 Earth masses indicated by the dashed line and initial total masses from $1.0\ M_{\rm J}$ to $1.5\ M_{\rm J}$ in steps of $0.1\ M_{\rm J}$. Crosses indicate the time when Roche-lobe overflow occurs. Each panel shows the results of (a) in situ formation scenario, and, (b) migration scenario. The range of the estimated age of Kepler-10b is shown by the gray zone. \label{k10b}}
\end{figure}

The results are shown in Fig. \ref{k10b}.
The maximum mass for Kepler-10b to be able to evolve from an initial gas giant planet into a rocky super Earth would be $ 1.1\ M_{\rm J}$.
This indicates that a Jupiter mass gas giant planet could evolve into a super Earth like Kepler-10b through thermal atmospheric mass loss and Roche-lobe overflow.
For Kepler-10b the maximum initial mass is smaller than that of CoRoT-7b because of the lower XUV flux level.
Kepler-10b orbits a hotter host star and has a slightly smaller orbital distance than CoRoT-7b.
The hotter environment makes atmospheric mass loss more intensive. 
However, the XUV luminosity has more dominant effect on the mass loss evolution and the lower XUV luminosity assumed for Kepler-10 results in the smaller maximum initial mass of Kepler-10b.

\section{Discussions}

\begin{figure}
  \includegraphics[width=0.99\columnwidth]{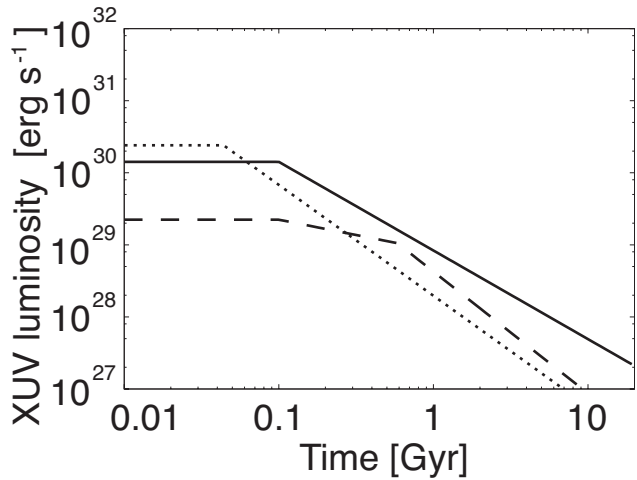}
\caption{Evolution models of stellar XUV (or X-ray) luminosity as a function of time: XUV scaling law of \citet{rib05}, assuming a saturation phase for the first $0.1\ {\rm Gyr}$ (solid line). X-ray scaling law of \citet{pen08}, assuming a saturation phase for the first $0.1\ {\rm Gyr}$ (dashed line). X-ray scaling law of \citet{zul12}, based on the result of \citet{gar11} (dotted line). \label{XUVmodels}}
\end{figure}

XUV flux dominates planetary mass loss.
Here we summarize the effect of different stellar XUV flux evolution models (Fig. \ref{XUVmodels}) and compare the results obtained.  
\citet{rib05} investigated the XUV luminosity of six G type stars, including the Sun. 
The stars are nearby solar proxies whose rotation periods, temperatures, luminosities, and metallicities are well-determined.
The range of the observed XUV is $1-1180\ {\rm \AA}$.
The data were obtained by an extrapolation for the range of $360-920\ {\rm \AA}$ where absorption by interstellar materials is strong.
This model was used in our calculations. 
To quantify the sensitivity of our results to different stellar XUV evolution models, we apply different XUV models and calculate the evolution of Kepler-10b.

The model by \citet{pen08} is based on the mean X-ray luminosities of G-type stars in  stellar clusters.
They employed mean values motivated by the fact that the stellar X-ray luminosities vary by an order of magnitude.
The work focused on X-ray luminosity ($1-200\ {\rm \AA}$) because of the interstellar absorption of EUV.
\citet{gar11} collected the mean X-ray luminosities of G-, K-, and M-type stars in stellar clusters.
\citet{zul12} derived an evolution model of the stellar X-ray luminosity (Eq. 23 in their paper) based on the result of \citet{gar11}.
The duration of the saturation phase is determined by connecting the two functions in Eq. 23 continuously.

Adopting the \citet{pen08} X-ray model for evolution models of Kepler-10b and maintaining the assumption of the $0.1\ {\rm Gyr}$ saturation phase \citep[following][]{lei11} lead to a maximum mass loss of $0.8\ M_{\rm J}$ for Kepler-10b for in situ scenario and migration scenario, changing the maximum mass loss by $-27\ \%$ for both scenarios.
Adopting the \citet{zul12} X-ray model, which has a $\sim 0.04\ {\rm Gyr}$ saturation phase, leads to a maximum mass loss of $1.1\ M_{\rm J}$ for in situ formation and $1.0\ M_{\rm J}$ for migration. 
These are changes of $0\ \%$  and $-9.1\ \%$ for in situ formation and migration scenario respectively, comparing to our model.
This analysis shows that the XUV evolution by Ribas et al (2005) provides a higher XUV environment and consequently higher planetary mass loss than models by \citet{pen08} and \citet{zul12} because EUV is not included in their power laws. 
Note that our results should be used as upper limits on planetary mass loss for Kepler-10b and CoRoT-7b.

The mass loss evolution of CoRoT-7b and Kepler-10b has been studied by several previous works.
\citet{val10} investigated the mass-loss evolution of CoRoT-7b and concluded that CoRoT-7b could be a remnant of a evaporated hot Jupiter.
\citet{jac10} studied the mass loss and the tidal evolution and concluded that the initial mass of CoRoT-7b does not exceed 200 Earth masses (2/3 Jupiter masses).
The differences of our study from these works are the updated XUV model based on the observation of X-ray of CoRoT-7 and the mass-loss model which takes Roche-lobe overflow into account, both of which result in the enhanced mass-loss.
The orbital evolution is not included our model.
The thermal atmospheric escape and Roche-lobe overflow might affect the orbital evolution as well as the tide.
\citet{lei11} investigated the mass loss evolution of CoRoT-7b and Kepler-10b and concluded that these planets are not a remnant of a hot Jupiter.
They used a scaling law for the evolution of the planetary radius.
On the other hand, our model solves the evolution by calculating the structure of the planet consistently and finds different results.

In this paper, we use the energy-limited formula (Eq. \ref{evo_escape}) to calculate the thermal atmospheric escape, \citep[following][]{val10,lei11}.
Thermal escape has a limiting mechanism, called the radiation-recombination limited escape, under extremely high XUV flux ($\sim 10^4\ {\rm erg\ cm^{-2}\ s^{-1}}$) condition \citep{mur09}.
The radiation-recombination limit mechanism is caused by the thermostat effect of Ly$\alpha$ cooling.
The XUV levels assumed in this work exceed that value in several cases.
This limiting mechanism would decrease the thermal mass loss rate and reduce the initial maximum planetary mass obtained in our simulations.
Exploring the effect on the planetary evolution is beyond the scope of this paper.
In addition, non-thermal escapes such as the interaction with the stellar wind \citep[e.g.,][]{Ekenback+2010} can affect the total mass of the planet in the long term evolution.



In our calculation, $L_{\rm int} = 10^{-5}\ L_{\odot}$ is assumed for initial intrinsic luminosity.
The effect of the initial intrinsic luminosity on the evolution can be understood by comparing the results of in situ formation and migration (Fig. \ref{Lint}).
The intrinsic luminosity at $0.01\ {\rm Gyr}$ (the time of initial mass loss) differs by an order of magnitude between the two scenarios but affects the maximum initial planetary mass to evolve into a rocky super Earth only slightly (Figs. \ref{c7b_1-2Gy}, \ref{c7b_2-3Gy}, and \ref{k10b}).

\section{Summary and Conclusion}

CoRoT-7b and Kepler-10b are hot super Earths, with measured masses and radii, and have potentially experienced intense mass loss through their evolution caused by stellar XUV irradiation.
We estimate the maximum mass of their initial planet by simulating the mass loss evolution of the hypothetical precursors using energy limited atmospheric escape and Roche-lobe overflow.
The results show that both planets could be remnants of Jupiter mass gas planets. 
The conclusion is independent of the assumptions on evolution model of the stellar XUV luminosity and on formation scenarios.
For such initial gas planets thermal atmospheric escape as well as Roche-lobe overflow determines the overall mass loss over the planet's evolution and its potential to evolve into a rocky super Earth in our model.
This is because CoRoT-7b and Kepler-10b have small Roche-lobe radii caused by their small masses and small orbital distances from the host stars.
The dependence on the formation scenario, which is interpreted as the difference of the intrinsic luminosity, is small.
Our results show that mass loss for close in planets mainly depends on the stellar XUV flux. 
This model can also be applied to other known extrasolar planets.

\section*{Acknowledgments}

The authors acknowledge support from DFG funding ENP Ka 3142/1-1. 
The authors acknowledge Prof. J. Zuluaga for fruitful discussions on the different XUV models and Dr. C. Mordasini for input for the equation of state of solar composition gas.
The authors appreciate the referee of this paper for many fruitful suggestions.


\label{lastpage}

\end{document}